\documentclass{aa} 
\usepackage{graphicx}
\usepackage{natbib}
\usepackage{txfonts}
\begin{document}
\title{Strong latitudinal shear in the shallow convection zone of a
  rapidly rotating A-star\thanks{Based on observations carried out at
    the European Southern Observatory, Paranal, 71.D-0127(A)}}


\author{A. Reiners\inst{1,2}
  \and
  M. H\"unsch\inst{3}
  \and
  M. Hempel\inst{4}
  \and
  J.H.M.M. Schmitt\inst{2}
}


\institute{
  Astronomy Department, University of California, Berkeley, CA 94720 \\ \email{areiners@astron.berkeley.edu}
  \and
  Hamburger Sternwarte, Universit\"at Hamburg, Gojenbergsweg 112, 21029 Hamburg, Germany \\ \email{jschmitt@hs.uni-hamburg.de}
  \and
  Institut f\"ur Theoretische Physik und Astrophysik, Leibnizstrasse 15, 24098 Kiel, Germany \\ \email{mhuensch@astrophysik.uni-kiel.de}
  \and
  Astrophysikalisches Institut Jena, Schillerg\"a\ss chen 2-3, 07745 Jena, Germany \\ \email{marc@astro.uni-jena.de}
}

\date{\today} 

\titlerunning{Strong latitudinal shear in IC4665~V102}

\abstract{We have derived the mean broadening profile of the star
  V\,102 in the region of the open cluster IC~4665 from high
  resolution spectroscopy. At a projected equatorial rotation velocity
  of $v\,\sin{i} = (105 \pm 12)$\,km\,s$^{-1}$ we find strong
  deviation from classical rotation. We discuss several scenarios, the
  most plausible being strong differential rotation in latitudinal
  direction. For this scenario we find a difference in angular
  velocity of $\Delta\Omega = 3.6 \pm 0.8$\,rad\,d$^{-1}$
  ($\Delta\Omega/\Omega = 0.42 \pm 0.09)$.  From the H$\alpha$ line we
  derive a spectral type of A9 and support photometric measurements
  classifying IC~4665 V\,102 as a non-member of IC~4665. At such early
  spectral type this is the strongest case of differential rotation
  observed so far.  Together with three similar stars, IC~4665 V\,102
  seems to form a new class of objects that exhibit extreme
  latitudinal shear in a very shallow convective envelope.

\keywords{stars: rotation -- line: profiles -- stars:
  individual(IC4665 V102/P35) -- clusters:individual(IC4665)} }

\maketitle
                                %

\section{Introduction}

The substantial difference between photospheres of solar-type stars
and A-type stars is the existence of a convective envelope.  Late-type
stars harbor convective envelopes where turbulent motions of
photospheric plasma can occur. In convective envelopes, differential
rotation can trigger a magnetic dynamo giving reason for the observed
plethora of activity phenomena. The onset of convection may be traced
by the onset of coronal X-ray emission around spectral type A7
\citep{Schmitt97}.  \cite{Gray89} searched for the onset of convection
analyzing line bisectors of slowly rotating stars.  In their targets a
bisector reversal was found around spectral type F0.  Stronger
asymmetries were found in the stars at the hot side of the boundary
indicating higher photospheric velocities.

While stellar rotation velocities have been measured for almost a
century, differential rotation is more difficult to detect
\citep[e.g.][and references therein]{Gray77, Gray98}.  Measurements of
differential rotation have been reported in stars rotating faster than
$v\,\sin{i} > 10$\,km\,s$^{-1}$ and covering all spectral types
harboring convective envelopes \citep{Reiners03b, Reiners03,
  Reiners04}. 

\section{Data}

\begin{figure*}
  \begin{center}
  \includegraphics[height=.9\hsize,angle=-90,clip=]{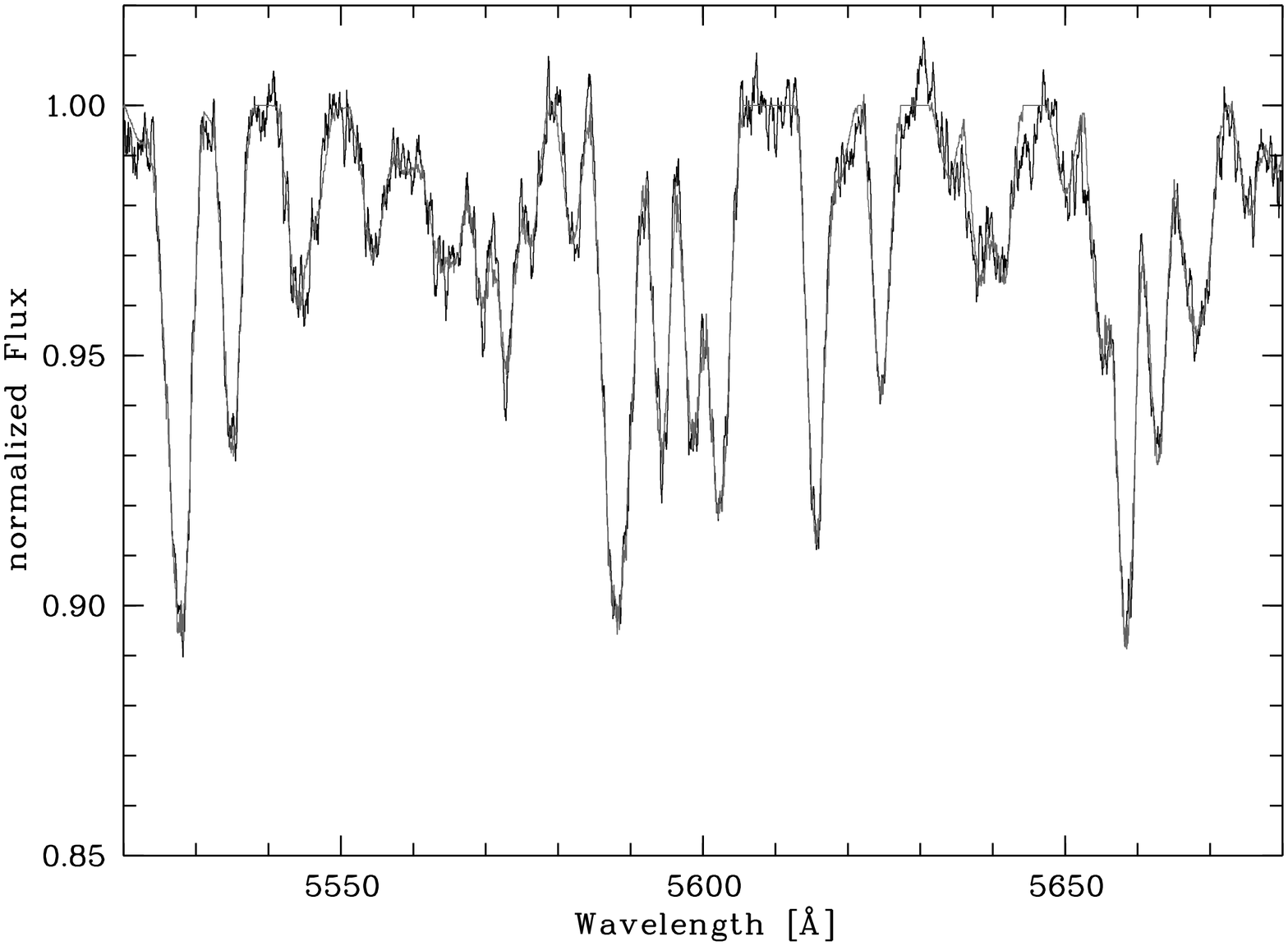}
  \end{center}
  \vspace{-5.5cm}

  \hspace{4.1cm}\includegraphics[height=4.5cm,width=3.3cm,angle=-90,clip=]{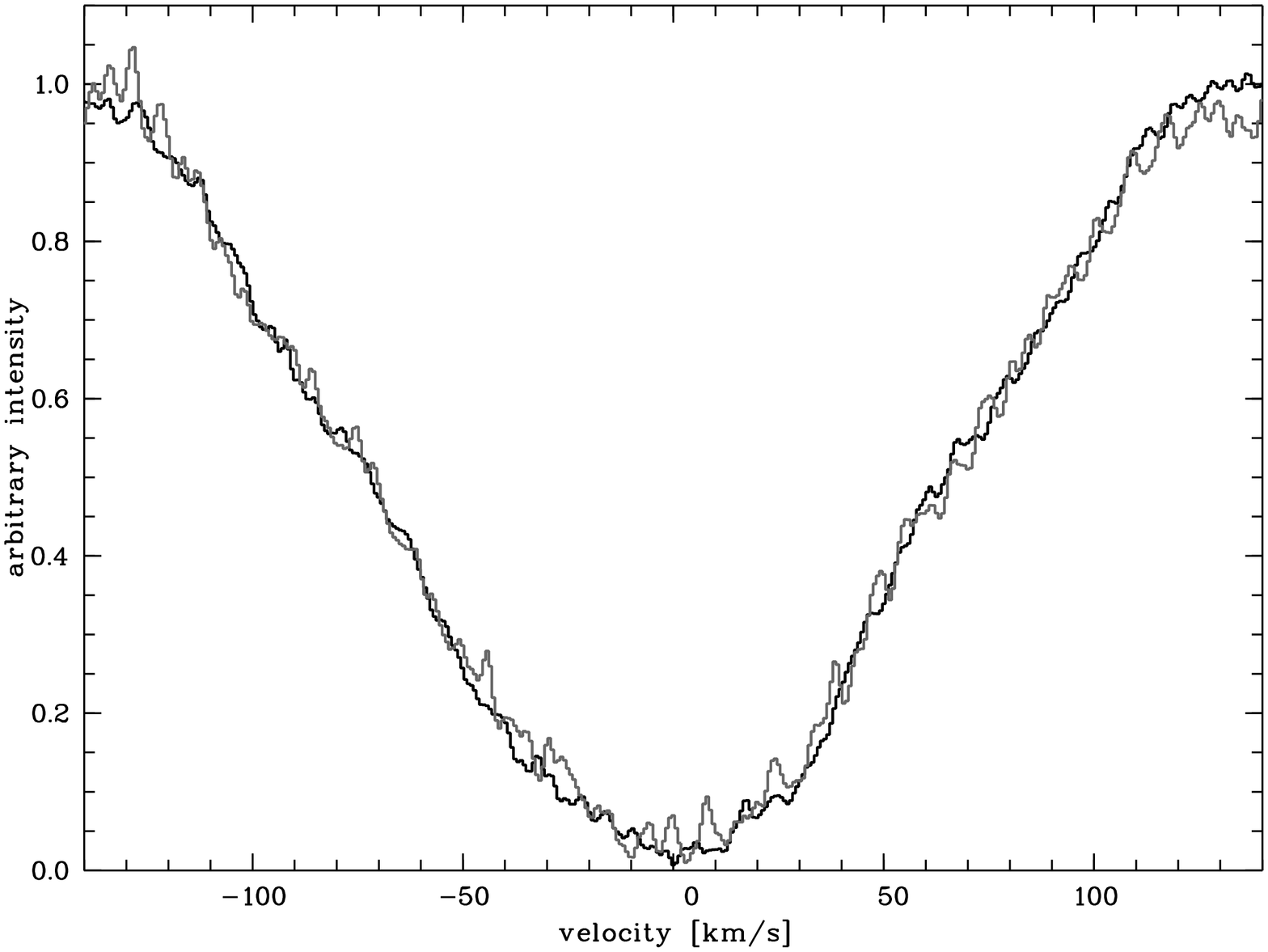}
  \vspace{1.5cm}
  \caption{\label{plot:Data}Data (black) and fit (grey) in a small part of the
    used wavelength region. In the inset the two broadening functions
    derived from the spectral regions 4950\,\AA\ -- 5550\,\AA\ (grey)
    and 5440\,\AA\ -- 5745\,\AA\ (black) are shown. Both broadening
    functions perfectly match one another. }
\end{figure*}

Our data of V\,102 have been taken on April 18th 2003 with the FLAMES
multi-object facility at ESO/VLT. Using the UVES spectrograph at a
resolution of 38\,000 we gained a signal-to-noise ratio of 150 in a
one hour exposure. The spectra were reduced using the UVES context
embedded in the ESO-MIDAS package. The observed spectral range covers
the region from 4800\,\AA\ to 6800\,\AA\ with a gap at 5800\,\AA. In a
second observation during the same night we collected a spectrum with
the FLAMES facility using the GIRAFFE spectrograph which covers the
Ca\textsc{ii}~H+K resonance lines. A small region of the UVES spectrum
is plotted in Fig.\,\ref{plot:Data} (black spectrum).

\section{V102 and its membership to IC~4665}
\label{sect:V102}

IC~4665 is a young \citep[50-70 Myr;][]{Prosser93, Giampapa98} open
cluster at a distance of roughly 350\,pc. The first photometric
measurements of V~102 were done by \cite{McCarthy69}. Although V~102
fitted onto the main sequence in their color-magnitude diagram, they
argued that V~102 is a ``less probable'' member of IC~4665 due to
their proper motion measurements.  More than twenty years later,
\cite{Prosser93} carried out a photometric study of IC~4665 and
derived more accurate proper motions.  From that he derived an $86\%$
membership probability for V~102.  However, he also carried out
CCD-photometry and found V~102 not to fit onto the cluster main
sequence according to its brightness in $V$. The results from
photometric measurements in V~102 are given in
Table\,\ref{tab:photometry}. The $V$-magnitudes derived by
\cite{McCarthy69} and by \cite{Prosser93} are significantly different
while $B$-magnitudes are identical and are also reproduced by the
value measured by \cite{Kislyuk99}. Since magnitudes in the $B$-filter
are identical in all three references we suspect that one of the
magnitudes in $V$ is incorrect, giving rise to the different values of
$B-V$.

\begin{figure}
  \centering
  \includegraphics[height=.8\hsize,angle=-90,clip=]{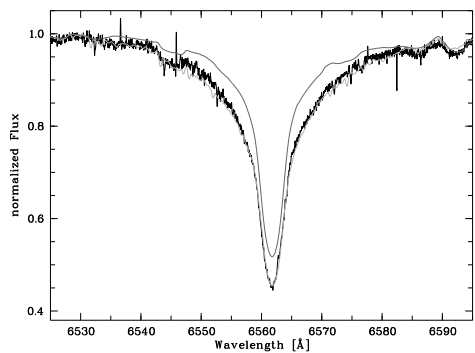}
  \caption{\label{plot:Halpha}H$\alpha$-line of V~102 (black).
    Overplotted are spectra of HD~115\,810 ($B-V=0.27$; light grey)
    and HD~120\,987 ($B-V=0.44$ spun up to $v\,\sin{i} \approx
    100$\,km\,s$^{-1}$; dark grey). The spectrum of HD~115\,810
    resembles the one of V~102 much better than the spectrum of
    HD\,120\,987 does.}
\end{figure}

The spectral types according to the $B-V$ measurements are either A9
or F5. In this spectral range the H$\alpha$-line is very sensitive to
temperature and we can deduce the spectral type from our data. The
H$\alpha$-line of V~102 is plotted in Fig.\,\ref{plot:Halpha} together
with spectra of HD~115\,810\footnote{taken with the FOCES
  spectrograph, DSAZ, Calar Alto} ($B-V=0.27,
v\,\sin{i}=100$\,km\,s$^{-1}$, light grey) and
HD~120\,987\footnote{taken with the FEROS spectrograph, ESO, La~Silla}
($B-V=0.44, v\,\sin{i}=8.5$\,km\,s$^{-1}$, dark grey) spun up to
$v\,\sin{i} \approx 100$\,km\,s$^{-1}$. From the excellent fit to the
H$\alpha$-line of HD~115\,810, we conclude that the color of V~102
must be similar to HD~115\,810. We thus adopt the value of $B-V=0.26$
given in \cite{Prosser93}, i.e., from photometry V~102 is likely to be
a background star and not a member of IC~4665 \citep{Prosser93}.

\begin{table}
  \caption{ \label{tab:photometry}Photometric data of IC~4665 V~102.   }
  \begin{tabular}{cccccr}
    \hline
    \hline
    \noalign{\smallskip}
    $V$ & $B$ & $B-V$ & $V-I$ & $B-R$ & designation\\
    \noalign{\smallskip}
    \hline
    \noalign{\smallskip}
    11.65 & 12.09 & 0.44 &&& V~102$^{\rm a}$\\
    11.82 & \emph{12.08}$^{\rm 1}$ & 0.26 & 0.41 && P~35$^{\rm b}$\\
    & 12.06 & & & 0.69 & 354\,699$^{\rm c}$\\
    \noalign{\smallskip}
    \hline
  \end{tabular}
  \begin{list}{}{}
  \item[$^{\rm 1}$]Calculated from $V$ and $B-V$, not given in the catalog
  \item[$^{\rm a}$]\cite{McCarthy69}
  \item[$^{\rm b}$]\cite{Prosser93}
  \item[$^{\rm c}$]\cite{Kislyuk99}
  \end{list}
\end{table}

\section{Broadening Function}

From our observations we derived the mean line broadening profile
following the strategy outlined in \cite{Reiners02a}. Normalization of
the spectrum was done by eye, even at rapid rotation the continuum is
still visible. Using absorption line wavelengths from the Vienna
Atomic Line Database \citep{VALD} as a reference, we iteratively
determine the shape of the broadening function and equivalent widths
of all incorporated lines.  To check against systematic errors we
performed the fit in two overlapping wavelength regions. The
wavelength region at 4950\,\AA\ -- 5550\,\AA\ incorporates 185 lines,
the second region at 5440\,\AA\ -- 5745\,\AA\ 137 lines.

\begin{figure*}
  \centering \mbox{
    \includegraphics[height=.45\hsize,width=5cm,angle=-90]{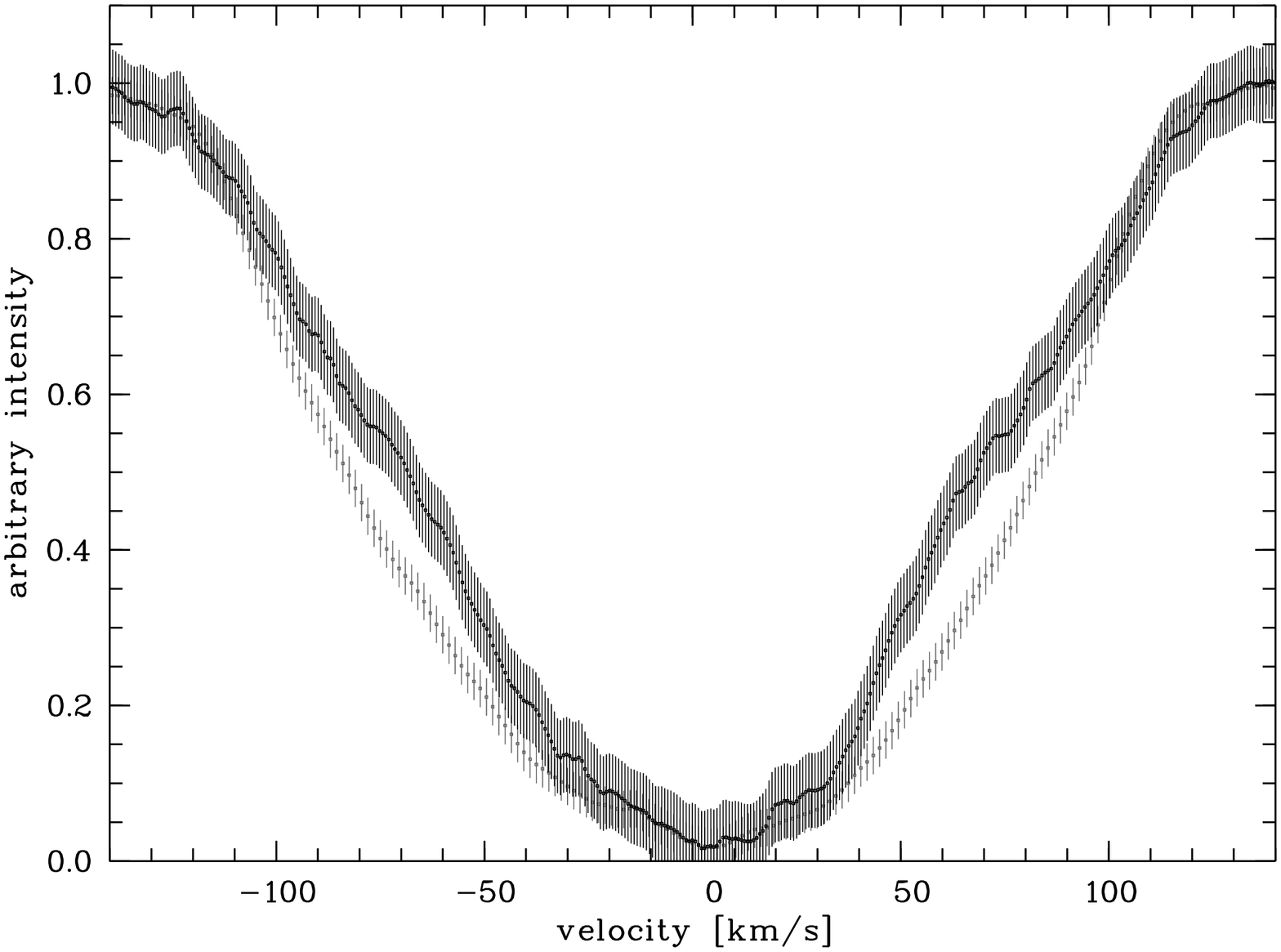}
    \includegraphics[height=.45\hsize,width=5cm,angle=-90]{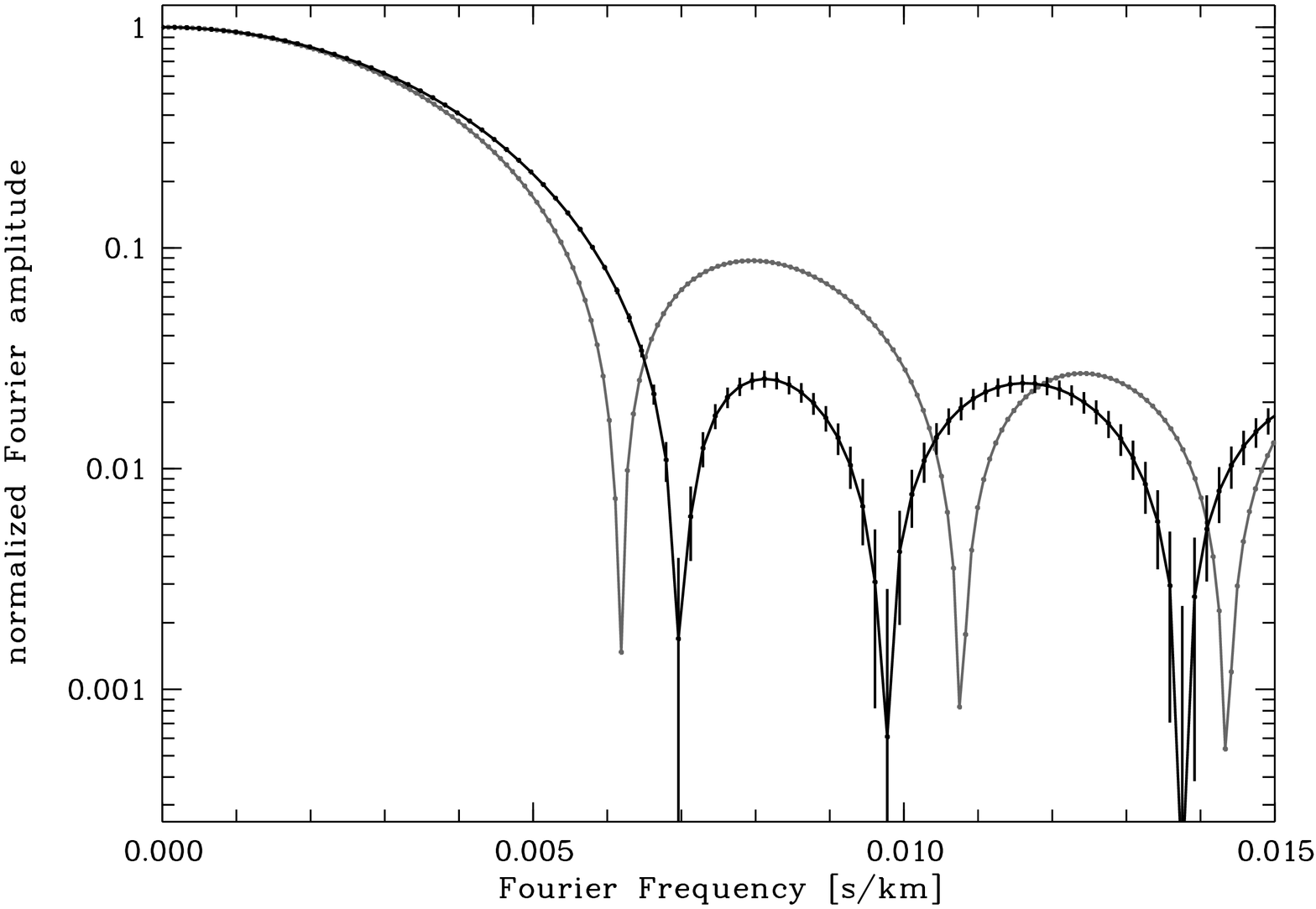} }
  \caption{\label{plot:LSF}Broadening functions of IC~4665 V~102 (black) 
    and HD~46\,273 (grey) in velocity (left panel) and Fourier space
    (right panel). The profile of HD~46\,273 is resembles classical
    rotation broadening. Significant differences are visible to the
    profile of IC~4665 V~102. Note especially the narrowed first
    sidelobe in the Fourier transform -- a typical signature for
    differential rotation with the equator rotating faster then the
    pole.}
\end{figure*}

We calculate the variances of the broadening function by taking into
account the uncertainties of the two nearest neighbors of each pixel.
We overplot in Fig.\,\ref{plot:Data} our final fit to the data in
grey.  The line shapes of all lines are well resembled within data
quality. In the inset of Fig.\,\ref{plot:Data} we compare the two
broadening functions independently derived from the two overlapping
wavelength regions mentioned above. Both broadening functions resemble
each other indicating indepence on the wavelength region; the
broadening function with its errors is plotted in black in the left
panel of Fig.\,\ref{plot:LSF}. Although every pixel was a free
parameter during the fit, the broadening profile is symmetric within
the uncertainties.

Quantitative conclusions can be drawn from the Fourier transform of
the broadening profile which is shown with its errors in the right
panel of Fig.\,\ref{plot:LSF}. In Fig.\,\ref{plot:LSF}, we also plot
the broadening profile of HD~46\,273, a star of similar spectral type
and $v\,\sin{i}$ (F2V, $v\,\sin{i} = 107$\,km\,s$^{-1}$). This profile
was achieved applying the same procedure to data obtained with the
FEROS spectrograph at ESO/La~Silla \citep{Reiners03}, it has a shape
as expected from classical solid body rotation.

\section{Interpretation}

The broadening function of V\,102 shows significant differences to
classical rotation broadening. Within the error bars a difference is
clearly visible between the profiles of V\,102 and HD~46\,273. At a
projected equatorial rotation velocity of $v\,\sin{i} =
105$\,km\,s$^{-1}$, the broadening functions are clearly rotation
dominated. Since the derived broadening functions are symmetric, it is
unlikely that they are dominated by starspots and we argue that the
profiles are due to the rotation law of the stellar surface.  In the
following, we will discuss the amount of differential rotation we
derive for V\,102 and the consequences for the surface shear.
However, we will also discuss several other scenarios that could be
reason for such a profile.

\begin{table}
  \caption{ \label{tab:summary}Results derived from the broadening function}
  \begin{tabular}{cccc}
    \hline
    \hline
    \noalign{\smallskip}
    $v\,\sin{i}$ & $\frac{q_2}{q_1}$ & $\alpha = \Delta\Omega/\Omega$  & $\Delta\Omega$ \\
    \noalign{\smallskip}
    \hline
    \noalign{\smallskip}
     $(105 \pm 12)$\,km\,s$^{-1}$ & $1.39 \pm 0.01$ & $0.42 \pm 0.09$ & $3.6 \pm 0.8$\,rad\,d$^{-1}$ \\
    \noalign{\smallskip}
    \hline
  \end{tabular}
\end{table}

\subsection{Differential rotation}

Convenient quantities characterizing stellar rotational broadening are
the first two zeros $q_{1}$ and $q_{2}$ of the broadening profile's
Fourier transform \citep[cf.][]{Reiners02a}. The value of $v\,\sin{i}$
can be derived from $q_1$. Furthermore, in a rigid rotator, the ratio
$q_{2}/q_{1}$ always is larger than 1.72. For V\,102 we derive a ratio
$q_{2}/q_{1} = (1.39 \pm 0.01) < 1.72$. This small ratio is visible in
the narrow first side-lobe of the Fourier transform compared to the
profile of HD~46\,273 in the right panel of Fig.\,\ref{plot:LSF}. To
quantify differential rotation we assume a rotation law parameterized
in analogy to the solar case:
\begin{equation}
  \label{DiffRotLaw}
  \Omega(l) = \Omega_{\rm Equator} (1 - \alpha \sin^2{l}),
\end{equation}
with $l$ heliographic latitude and $\Omega_{\rm Equator}$ angular
rotation speed at the equator.  Differential rotation can be described
by the parameter $\alpha$, which is the difference of the equatorial
and polar rotation velocities relative to the equatorial velocity.

As shown in \cite{Reiners03b}, this parameter $\alpha$ can be
calculated from the observed ratio $q_{2}/q_{1}$, specifically
$q_{2}/q_{1} \propto \alpha/\sqrt{\sin{i}}$, with $i$ the inclination
angle. For the case of V~102 we derive $\alpha = 0.42 \pm 0.09$, and
with a radius of $R~=~1.5$\,R$_{\rm \sun}$ the difference between
polar and equatorial angular velocities is
$\Delta~=~3.6~\pm~0.8$\,rad\,d$^{-1}$.

Such strong a shear in a fast rotator is expected to maintain a strong
magnetic dynamo and we would expect significant X-ray flux from V~102.
\cite{Giampapa98} carried out X-ray observations of IC~4665 with no
detection of X-ray flux from V~102. However, for a distance of
$d=350$pc, we estimate an upper limit of X-ray luminosity $L_{\rm
  X}\la3.4\,10^{29}$\,erg\,s$^{-1}$. This is about the typical X-ray
luminosity detected in that spectral range \citep{Pallavicini81}.
Since V~102 is likely to be a background object, the X-ray
observations do not contradict the picture of a strong magnetic dynamo
being driven by rapid differential rotation.

\subsection{Alternative scenarios}

Although we believe that differential rotation is the most probable
interpretation of the derived broadening profile, other mechanisms can
be potential reasons for the deviations from classical rotational
broadening as well. We discuss three alternative scenarios.

\subsubsection{Low inclination}

Very rapidly rotating stars become gravity darkened, i.e., they become
cooler at the equator. At (unprojected) rotational velocities larger
than 150\,km\,s$^{-1}$, the shape of the broadening profile, and
especially the ratio $q_{2}/q_{1}$, is affected by gravity darkening
\citep[cf.][]{Reiners03a}. From the models given in \cite{Reiners03a}
we can calculate the equatorial velocity needed for our observed ratio
of $q_{2}/q_{1} = 1.39$; we get a value of $v_{\rm{rot}} =
800$\,km\,s$^{-1}$, which is far beyond breakup velocity. Thus we can
rule out gravity darkening as the main reason for the observed
profile.

\subsubsection{Bright Polar Spot}

The surface of V\,102 may be covered with spots. If the star were
magnetically active, one would expect cool spots covering the star.
Cool spots generate characteristic signatures in absorption profiles.
Since the derived broadening profile is symmetric, the surface
coverage of spots must also be symmetric to a high degree. One
configuration that would satisfy these requirements is a polar spot.
We examined the influence of spots on the broadening profile
\citep[cf.][]{Reiners02b} and found that cool polar spots do change
the profile shape and the ratio $q_{2}/q_{1}$ but in the opposite
direction; $q_{2}/q_{1}$ becomes larger. A profile shape as the one
observed could only be due to a \emph{hot} polar spot. From our models
we estimate that a polar cap with a radius larger than $30^{\circ}$ at
a temperature about $1000$\,K hotter than the rest of the star is
needed to produce such a profile shape. This could be similar to what
is observed in T~Tauri stars accreting material at the poles, and one
obvious sign of such accretion would be H$\alpha$ emission. In V~102,
no sign of H$\alpha$ emission can be found (cp.
Fig.\,\ref{plot:Halpha}).

We emphasize that this may nevertheless be a scenario alternative to
differential rotation, but we see no physical connection to any
phenomenon observed on other stars.

\subsubsection{Composite spectrum}

The derived broadening profile could tentatively also be a composition
of two spectra from rigidly rotating stars. Especially the curvature
at around $\pm 70$\,km\,s$^{-1}$ could be interpreted as an indication
for that scenario (although it can also be explained by differential
rotation). The profile could then be interpreted by a spectrum of the
first star rotating at $v\,\sin{i} \approx 110$\,km\,s$^{-1}$,
superimposed by a spectrum of a second object rotating at $v\,\sin{i}
\approx 70$\,km\,s$^{-1}$. Thus both stars would have either
significantly different rotation velocities or were seen under
different angles of inclination, i.e., a hypothetical double system
could not be corotating.  Since we derived identical profiles from two
different wavelength regions covering several hundred \AA ngstr\"oms,
both stars had to be of similar spectral type, otherwise the line
profile would change at different wavelengths. In order to resemble
the derived profile, both stars must also contribute approximately the
same amount of flux. Given similar spectral types of the components,
that implies that both stars are at the same distance and are probably
physically bound. The double system would then emit roughly twice the
flux a single star would, i.e., the real magnitude of each member
would be 0.75\,mag higher than the measured ones putting V\,102 even
further away from the main sequence in the color-magnitude diagram.

In this scenario, V\,102 would be a physically bound system consisting
of two stars of similar spectral type and different rotation
velocities located far outside the cluster but comoving with the
cluster.  We would expect such a spectrum to be variable. Although we
consider this scenario very unlikely, we aim to reobserve the star to
check the profile for variability.

\section{Conclusions}

We derived the mean line broadening profile of V~102, a background
star of spectral type A9 in the region of the young open cluster
IC~4665. From the broadening profile we determined a projected
rotation velocity of $v\,\sin{i} = 105 \pm 12$\,km\,s$^{-1}$. The
broadening profile significantly differs from classical rotation
broadening and we discuss several scenarios for such a profile. The
most plausible scenario is that V~102 does not rotate as a solid body
but that its equator rotates at a higher velocity than the pole.  For
that case we derive a difference in polar and equatorial angular
velocities of $\Delta\Omega~=~3.6~\pm~0.8$\,rad\,d$^{-1}$, or
$\Delta\Omega/\Omega = 0.42 \pm 0.09$.

This is so far the strongest observed case of differential rotation.
The equator of V~102 laps the polar region once every 40~hours,
compared to roughly 120~days on the Sun. Three other stars with strong
differential rotation of the order of $\Delta\Omega/\Omega = 0.3$ have
been reported in \cite{Reiners04}, they have colors of $B-V=0.29$ or
$0.30$.  Our result supports evidence for differential rotation being
strongest in a class of rapidly rotating objects with very shallow
convection zones.

\begin{acknowledgements}
  A.R. has received research funding from the European Commission's
  Sixth Framework Programme as an Outgoing International Fellow
  (MOIF-CT-2004-002544).
\end{acknowledgements}


\begin{thebibliography}{} 
  
  
\bibitem[Cox, 2000]{Allen} Cox, A.N., 2000, Allen's astrophysical
  quantities, 4th ed., AIP Press, Springer, New York,~Editedy by
  Arthur N.~Cox

  
\bibitem[Giampapa et al., 1998]{Giampapa98} Giampapa, M.S., Prosser,
  C.F.~\& Fleming, T.A., 1998, ApJ, 501, 624
  
\bibitem[Gray, 1977]{Gray77} Gray, D.~F., 1977, \apj, 211, 198

\bibitem[Gray \& Nagel, 1989]{Gray89} Gray, D.~F., \& Nagel, T., 1989,
  \apj, 341, 421
  
\bibitem[Gray, 1998]{Gray98} Gray, D.~F., 1998, ASP Conf. Ser. 154,
  Tenth Cambridge Workshop on Cool Stars, Stellar Systems and the Sun,
  p.~193

\bibitem[Kislyuk et al., 1999]{Kislyuk99} Kislyuk, V., Yatsenko, A.,
  Ivanov, G., Pakuliak, L., \& Sergeeva, T., 1999, The FON
  Astrographic Catalogue, Main Astronomical Observatory of NASU
 
\bibitem[Kupka et al., 1999]{VALD} Kupka, F., Piskunov, N.E.,
  Ryabchikova, T.A., Stempels, H.C.~\& Weiss, W.W., 1999, A\&AS, 138,
  119
  
\bibitem[McCarthy \& O'Sullivan, 1969]{McCarthy69} McCarthy, M.F.~\&
  O'Sullivan, S.J., Ric. Astron., 7, 483
  
\bibitem[Pallavicini et al., 1981]{Pallavicini81} Pallavicini, R.,
  Golub., L., Rosner, R., Vaiana, G.S., Ayres, T.~\& Linsky, J.L.,
  1981, \apj, 248, 279

\bibitem[Prosser, 1993]{Prosser93} Prosser, C.F., 1993, AJ, 105, 1441
  
\bibitem[Reiners, 2003]{Reiners03a} Reiners, A., 2003, \aap, 408, 707
  
\bibitem[Reiners \& Schmitt, 2003a]{Reiners03b} Reiners, A.~\& Schmitt,
  J.H.M.M., 2003a, \aap, 398, 647
  
\bibitem[Reiners \& Schmitt, 2003b]{Reiners03} Reiners, A.~\& Schmitt,
  J.H.M.M., 2003b, \aap, 412, 813
  
\bibitem[Reiners \& Schmitt(2002a)]{Reiners02a} Reiners, A.~\&
  Schmitt, J.H.M.M.\ 2002a, \aap, 384, 155
  
\bibitem[Reiners \& Schmitt(2002b)]{Reiners02b} Reiners, A.~\&
  Schmitt, J.H.M.M.\ 2002b, \aap, 388, 1120
  
\bibitem[Reiners \& Royer, 2004]{Reiners04} Reiners, A.~\& Royer, F.,
  2004, A\&A, 415, 325
  
\bibitem[Schmitt, 1997]{Schmitt97} Schmitt, J.H.M.M., 1997, A\&A, 318,
  215
   
\end{thebibliography}
\end{document}